\documentclass[%
 reprint,
 amsmath,amssymb,
 aps
 prl
]{revtex4-1}

\usepackage[english]{babel}

\makeatletter
\def\bbl@set@language#1{%
  \edef\languagename{%
    \ifnum\escapechar=\expandafter`\string#1\@empty
    \else\string#1\@empty\fi}%
  \@ifundefined{babel@language@alias@\languagename}{}{%
    \edef\languagename{\@nameuse{babel@language@alias@\languagename}}%
  }%
  \select@language{\languagename}%
  \expandafter\ifx\csname date\languagename\endcsname\relax\else
    \if@filesw
      \protected@write\@auxout{}{\string\select@language{\languagename}}%
      \bbl@for\bbl@tempa\BabelContentsFiles{%
        \addtocontents{\bbl@tempa}{\xstring\select@language{\languagename}}}%
      \bbl@usehooks{write}{}%
    \fi
  \fi}
\newcommand{\DeclareLanguageAlias}[2]{%
  \global\@namedef{babel@language@alias@#1}{#2}%
}
\makeatother

\DeclareLanguageAlias{en}{english}

\usepackage{graphicx}
\usepackage{amsmath}
\usepackage{array}
\usepackage{amssymb}
\usepackage{dcolumn}
\usepackage{bm}
\let\savecorresponds\corresponds
\let\corresponds\relax
\usepackage{mathabx}
\usepackage{enumitem} 
\usepackage{tabularx}
\let\corresponds\savecorresponds

\def\vec#1{\boldsymbol{#1}}

\def\pd2v#1#2#3{\frac{\partial^2 #1}{\partial #2 \partial #3}}

\def \vec#1{\mathbf{#1}}
\def \2x2mat#1#2#3#4{
\left( \begin{array}{cc}
#1 &  #2 \\  #3 &  #4
\end{array} \right)
}

\begin{document}

\preprint{APS/123-QED}

\title{Spatially-entangled Photon-pairs Generation Using Partial Spatially Coherent Pump Beam}

\author{Hugo Defienne}
\email{hugo.defienne@gmail.com}
\author{Sylvain Gigan}%
\affiliation{%
Laboratoire Kastler Brossel,ENS-Université PSL, CNRS, Sorbonne Universite,\\
College de France, 24 rue Lhomond, 75005 Paris, France
}%

\date{\today}

\begin{abstract}
We demonstrate experimental generation of spatially-entangled photon-pairs by spontaneous parametric down conversion (SPDC) using a partial spatially coherent pump beam. By varying the spatial coherence of the pump, we show its influence on the downconverted photon's spatial correlations and on their degree of entanglement, in excellent agreement with theory. We then exploit this property to produce pairs of photons with a specific degree of entanglement by tailoring of the pump coherence length. This work thus unravels the fundamental transfer of coherence occuring in SPDC processes, and  provides a simple experimental scheme to generate photon-pairs with a well-defined degree of spatial entanglement, which may be useful for quantum communication and information processing. 
\end{abstract}

\maketitle

Quantum entanglement is considered as one of the most powerful resource for quantum information. In this respect, pairs of photons are the simplest system showing genuine quantum entanglement in all their degrees of freedom: spatial, spectral and polarization~\cite{brendel_pulsed_1999,kwiat_new_1995,howell_realization_2004}. Most of the fundamental experiments and related applications are implemented using polarization-entangled photons. Examples range from the first test of Bell's inequality~\cite{aspect_experimental_1982} to the recent development of long-distance quantum communication systems~\cite{liao_satellite--ground_2017}. In the last years, there has been renewed interest in continuous variable entanglement between transverse position and momentum of photon-pairs~\cite{walborn_spatial_2010}. Indeed, their infinite-dimensional Hilbert space holds high potential for developing powerful information processing algorithms~\cite{tasca_continuous-variable_2011} and secured cryptography protocols~\cite{walborn_quantum_2006}. Furthermore, spatially-entangled photon-pairs sources are at the basis of many quantum imaging approaches, including ghost imaging~\cite{pittman_optical_1995}, sub-shot-noise~\cite{brida_experimental_2010} and sub-Rayleigh imaging~\cite{xu_experimental_2015}. All these quantum applications crucially rely on properties of the down-converted photons. In this respect, their degree of entanglement is a fundamental parameter that generally defines the power of the quantum-based technique. As concrete examples, it sets the information bound in high-dimensional quantum communication systems~\cite{dixon_quantum_2012} and the spatial resolution in certain quantum imaging scheme~\cite{reichert_biphoton_2017}. However, most apparatus used to produce entangled pairs are not flexible and adapting pairs characteristics to specific use is generally a challenging task. In this work, we propose an novel experimental approach based on spontaneous parametric down conversion (SPDC) with a partial spatially coherent pump beam  to produce entangled photon-pairs with tunable degree of spatial entanglement.

\begin{figure}
\includegraphics[width=1 \columnwidth]{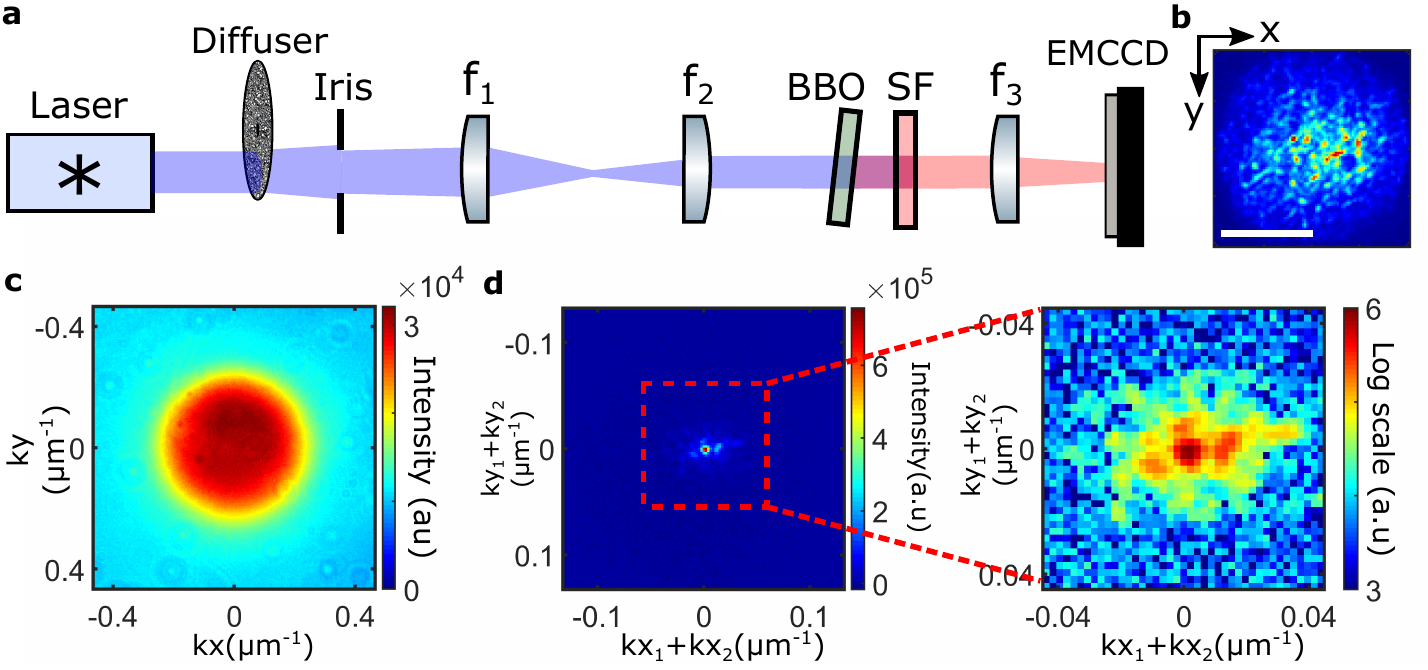}
\caption{\label{Figure1}  \textbf{(a)} Light emitted by a diode laser ($\lambda_p=405nm$) is scattered by a static thin diffuser (plastic sleeve) and illuminates a non-linear crystal of $\beta$-Baryum Borate (BBO) to produce spatially-entangled pairs of photons by type I SPDC. Spectral filters at $810 \pm 10$nm select near-degenerate photons. Lenses $f_1 =150$mm and $f_2 = 200$mm images an iris onto the cristal surface. When the diffuser is maintained fixed, the crystal is thus illuminated by a static speckle pattern \textbf{(b)}.  White scale bar corresponds to $ 700 \mu m$. Momenta of photons are imaged onto an EMCCD camera by imaging the far-field via a $f_3 = 40$mm lens, and a direct intensity image \textbf{(c)} is recorded by accumulating photons onto an EMCCD camera sensor. Sum-coordinate projection of the joint probability distribution  of photon-pairs \textbf{(d)} shows a coincidence speckle pattern that reveal the transfer of coherence between the pump and the down-converted fields.}
\end{figure}

\begin{figure*}
\includegraphics[width=1 \textwidth]{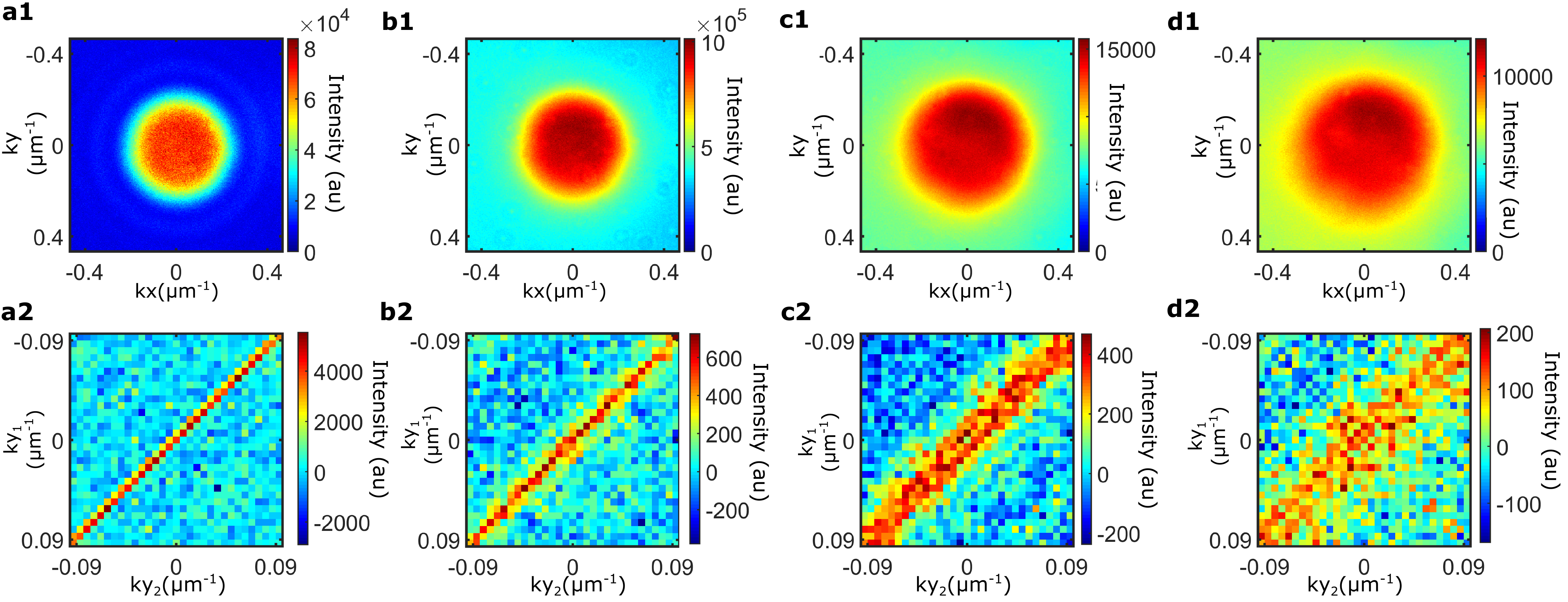}
\caption{\label{Figure2} Without diffuser, the direct intensity image \textbf{(a1)} shows a well defined disk and the $X_+$-coordinate projection of $\Gamma$ \textbf{(a2)} shows a strong anti-diagonal. An element $(ky_1,ky_2)$ of the $X_+$-coordinate projection corresponds to the joint probability of detecting one photon at $\vec{k_1} = (k_{x_1},k_{y_1})$, with no constraints on $k_{x_1}$, together with the second photon at $\vec{k_2} = (-k_{x_1},k_{y_2})$.  The strong anti-diagonal is a signature of momentum conservation between photons produced by SPDC with a fully coherent collimated pump beam. When a rotating random diffuser composed by one layer of plastic sleeve is introduced in the apparatus, edges of the direct intensity disk gets blurred \textbf{(b1)} and the width of the anti-diagonal on the $X_+$-coordinate projection increases \textbf{(b2)}. When one photon is detected at $\vec{k}$, its twin has now a high probability to arrive on an area that spreads around $-\vec{k}$. This area gets broaden when the coherence length of the pump is decreased by using rougher random diffusers, as shown on direct images \textbf{(c1)} and \textbf{(d1)} and $X_+$-projections \textbf{(c2)} and \textbf{(d2)} measured using respectively two layers of plastic sleeve and three layers.}
\end{figure*}

SPDC is the most popular technique to produce spatially-entangled photon-pairs. In its conventional form, a coherent Gaussian beam of light (i.e. the pump beam) illuminates a non-linear crystal ($\chi^2$ non-linearity) that produces pairs of photons in accordance with energy and momentum conservation~\citep{hong_theory_1985}. Properties of down-converted photons, including their degree of entanglement, are set by the crystal parameters and the pump beam properties~\cite{rubin_transverse_1996,souto_ribeiro_partial_1997,joobeur_coherence_1996,fonseca_transverse_1999,saleh_wolf_2005}. 
During this process, coherence properties of the pump beam get entirely transferred to those of the two photon-field~\cite{monken_transfer_1998,kulkarni_transfer_2017,ismail_polarization-entangled_2017}. Interestingly, none of these experimental studies consider the use of a non-perfectly spatially coherent pump beam to produce photon-pairs, with the notable exception of the recent work of Y. Ismael et al.~\cite{ismail_polarization-entangled_2017} that investigates polarization-entanglement between photons. Theoretically, the link between spatial coherence properties of the pump and the degree of entanglement of the down-converted field has been precisely established in~\cite{jha_spatial_2010, olvera_two_2015, giese_influence_2018}. In this work, we first investigate experimentally the influence of the pump spatial coherence on the correlation properties of the spatially-entangled photon pairs. We then demonstrate the dependency of the degree of entanglement, characterized by the Schmidt number~\cite{fedorov_gaussian_2009}, with the coherence of the pump. Finally, we exploit this effect to generate photon-pairs with a well-defined degree of entanglement by manipulating the transverse coherence length of the pump.

Figure~\ref{Figure1}.a shows the apparatus used to produce spatially entangled photon-pairs. A partially coherent beam of light is generated by intercepting the propagation path of a continuous-wavelength ($405$nm) Gaussian laser beam with a (rotating or not) random diffuser (plastic sleeve). Blue photons interact with a tilted non-linear crystal of $\beta$-baryum borate (BBO) to produce infrared pairs of photons by type I SPDC. At the output of the crystal, transverse momentum $\vec{k}$ of photons is mapped onto pixels of an electron multiplied charge coupled device (EMCCD) camera by a Fourier-lens imaging system ($f_3$). When the diffuser is maintained static, the crystal is illuminated by a speckle pattern (Figure~\ref{Figure1}.b). A direct intensity image (Figure~\ref{Figure1}.c) is acquired by photons accumulation on the camera sensor and shows an homogeneous structure, very similar to the one observed without diffuser (Figure~\ref{Figure2}.a1). However, when measuring the joint probability ditribution $\Gamma$ with the EMCCD camera~\cite{reichert_massively_2018,defienne_general_2018}, its projection along the sum-coordinate diagonal shows a central peak surrounded by a speckle pattern (Figure~\ref{Figure1}.c). The sum-coordinate projection represents the probability of detecting the two photons with symmetric momentum relative to their mean $\vec{k_1}+\vec{k_2}$~\cite{moreau_realization_2012,tasca_imaging_2012} (see~\cite{supmat} section 4). The presence of this speckle together with the absence of any spatial structure in the direct intensity image demonstrates that first-order spatial coherence of the pump field (i.e. intensity speckle pattern) gets entirely transferred to second-order coherence of the down-converted field (i.e. coincidence speckle pattern). 

\begin{figure}
\includegraphics[width=1 \columnwidth]{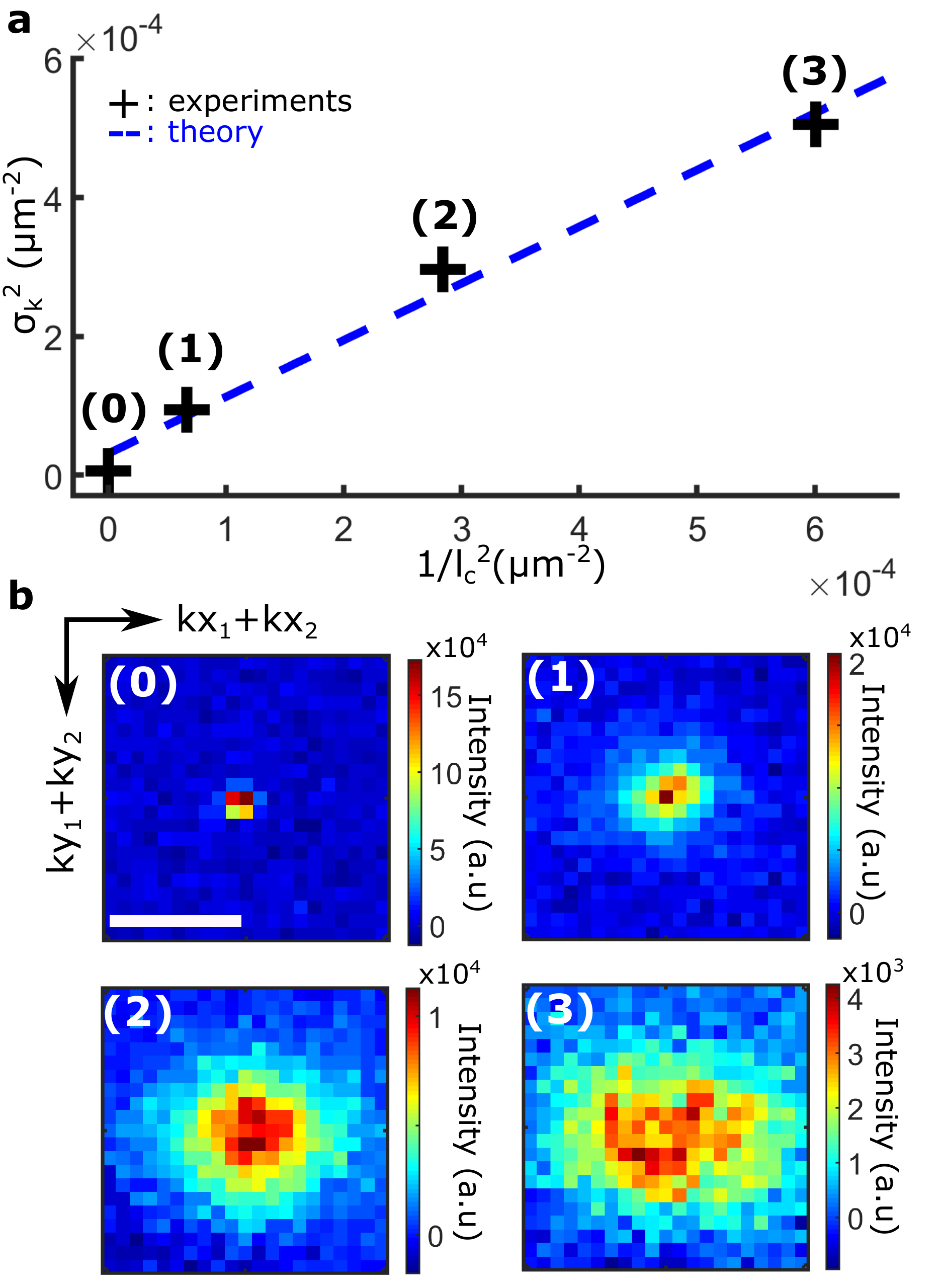}
\caption{\label{Figure3}  \textbf{(a)} Momentum-correlation width $\sigma_k$ is represented in function of coherence length of the pump $\ell_c$. Values on the graph correspond to four different measurements performed (0) without diffuser, (1) with one layer of plastic sleeve, (2) two layers and (3) three layers. Linear regression fits experimental values with a determination coefficient of $0.98$ and returns a slope value of $0.82 \pm 0.3$, in accordance with equation~\ref{eq2}. Values of $\sigma_k$ are estimated in each cases by projecting $\Gamma$ along the sum-coordinates diagonal \textbf{(b)} and measuring the width of the central spot using a Gaussian model~\cite{fedorov_gaussian_2009}.  $\ell_c$ values are estimated with a technique described in~\cite{supmat} section 4. White scale bar corresponds to $ 0.05 \mu m^{-1}$.    }
\end{figure}

As a consequence, spatial incoherence properties of the pump must be retrieved in the momentum correlations of the pairs. When the diffuser is rotated faster than the camera integration time, the pump acts as a partial spatially coherent beam. Using a Gaussian-Schell model for the pump beam~\cite{mandel_coherence_1965} and a Gaussian approximation for the down converted field~\cite{fedorov_gaussian_2009} (see~\cite{supmat} section 1), $\Gamma$ is written as 
\begin{equation}
\label{eq1}
\Gamma(\vec{k_1},\vec{k_2}) \sim \exp \left(- \frac{ \sigma_r^2  |\vec{k_1}-\vec{k_2}|^2}{2} \right) \exp \left(- \frac{|\vec{k_1}+\vec{k_2}|^2}{2 \sigma_k^2} \right)
\end{equation}
The position-correlation width $\sigma_r$ only depends on the crystal length $L$ and the pump frequency $\lambda_p$ as $\sigma_r = \sqrt{\alpha L \lambda_p/(2 \pi)}$ ($\alpha = 0.455$~\cite{chan_transverse_2007}). The momentum-correlation width $\sigma_k$ depends on the pump beam waist $\omega$ and its correlation length $\ell_c$ as
\begin{equation}
\label{eq2}
\sigma_k  = \sqrt{\frac{1}{\ell_c^{2}} + \frac{1}{4 \omega^{2}}}
\end{equation}
For a given crystal, varying the coherence properties of the pump beam (i.e. waist and correlation length) modifies the spatial structure of the two-photon wave function and its associated joint probability distribution. In particular, decreasing the correlation length at fixed waist induces an increase of the momentum-correlation width: when one photon of a pair is detected at $\vec{k}$, the area of maximum probability detection for its twin is centered at $-\vec{k}$ and spreads as $\sigma_k^2 \sim \ell_c^{-2}$. This effect is shown in Figure~\ref{Figure2}. For a perfectly coherent pump beam (no diffuser), the direct intensity image (Figure~\ref{Figure2}.a1) shows a well-defined homogeneous disk and the $X_+$-projection of $\Gamma$ (Figure~\ref{Figure2}.a2) shows a strong anti-diagonal. The $X_+$-projection image represents the joint probability of detecting one photon with momentum $k_{y_1}$  ($k_{x_1}$ can take any possible values) and its twin with momentum $k_{y_2}$ and $k_{x_2}=-k_{x_1}$ (see~\cite{supmat} section 4). Such strong anti-diagonal is a clear signature of transverse momentum conservation in SPDC using a collimated pump beam. When a rotating diffuser is used (single layer of plastic sleeve), the pump beam becomes partially coherent which results in a blurring of the edges of the direct intensity disk (Figure~\ref{Figure2}.b1) and an increase of the diagonal width in the $X_+$-coordinate projection (Figure~\ref{Figure2}.b2). Broadening of momentum correlations with the decrease of pump spatial coherence shows very well when using rougher diffusers, respectively made by superimposing two layers of plastic sleeves (Figure~\ref{Figure2}.c1 and c2) and three layers (Figure~\ref{Figure2}.d1 and d2). A quantitative analysis of this effect is provided in Figure~\ref{Figure3}. On the one hand, values of $\sigma_k$ are determined by fitting sum-coordinate projection of $\Gamma$ (Figure~\ref{Figure3}.b) by a Gaussian model~\cite{fedorov_gaussian_2009}. On the other hand, values of $\ell_c$ are measured by removing the crystal and Fourier-imaging the pump beam directly onto the camera (see~\cite{supmat} section 3). The linear regression of $\sigma_k^2 = f(1/\ell_c^2)$ (Figure~\ref{Figure3}.a) returns a slope value of $0.82 \pm 0.3 $ with a determination coefficient of $0.98$. This result is in very good accordance with equation~\ref{eq2} and shows the relevance of the theoretical model~\cite{jha_spatial_2010,giese_influence_2018}.

\begin{table*}
\caption{\label{table1} Values of $\sigma_k$ and $\sigma_k$ are listed in function of the coherence length $\ell_c$ of the pump beam. The Schmidt number $K$ (exp.) is calculated using the formula $K = 1/4 \left[ 1/(\sigma_r \sigma_k)+\sigma_r \sigma_k\right]^2$. Because $\sigma_r$ does not depend on the coherence property of the source, $K$ decreases with the diminution of the coherence length $\ell_c$. Theoretical values of $K$ [$K$ (theory)] are calculated from crystal parameters $L\approx 0.9mm$ and pump properties $\omega \approx 89 \mu$m and $\lambda_p \approx 405nm$ using equation~\ref{eq3}. Despite many approximations used and several sources of experimental uncertainties, the theoretical model is in good accordance in order of magnitude with experimental measurements. }
\begin{ruledtabular}
\begin{tabular}{ccccc}
 Correlation length ($\mu$m) & $\sigma_k (rad.mm^{-1} ) $ & $\sigma_r (\mu m)$ & $K$ (exp.) &  $K$ (theory) \\
 \hline
$+\infty$ & $2.4 \pm 0.1$ & $7.9 \pm 0.3$ & $727 \pm 74$ & $591$ \\
122 & $9.7 \pm 0.3$ & $8.4 \pm 0.2$ & $38 \pm 4$ & $115$ \\
59 & $17.2 \pm 0.3$ & $7.1 \pm 0.1$ & $17 \pm 1$ & $32$ \\
41 & $22.5 \pm 0.5$ & $7.1 \pm 0.2$ & $10 \pm 1$ & $16$ \\
\end{tabular}
\end{ruledtabular}
\end{table*}

Not only does partial coherence influence momentum correlations between pairs, but it also modifies their degree of entanglement. An universal metric to quantify it is the Schmidt number $K$, that is directly related to the non-separability of the two-photon state~\cite{law_analysis_2004}. Experimentally, $K$ is estimated from measurements of $\sigma_k$ and $\sigma_r$ using the formula $K = 1/4 \left[ 1/(\sigma_r \sigma_k)+\sigma_r \sigma_k\right]^2$~\cite{fedorov_schmidt_2015}. While $\sigma_k$ is determined using the apparatus described previously (Figure~\ref{Figure1}), values of $\sigma_r$ are measured using a different experimental configuration in which the output surface of the crystal is imaged onto the EMCCD camera (see~\cite{supmat} section 2). As reported in Table~\ref{table1}, $\sigma_r$ is constant for all diffusers and does not depend on the pump coherence properties. In consequence, the measured degree of entanglement $K$ [$K$ (exp.) in Table~\ref{table1}] decreases with the reduction of the correlation length $\ell_c$. As a comparison, values of $K$ [$K$ (theory) in Table~\ref{table1}] are calculated directly from crystal and pump properties using the theoritical model (equation~\ref{eq1}) 
\begin{equation}
\label{eq3}
K = \frac{1}{4} \left[ \frac{2 \omega l_c \sqrt{2 \pi}}{ \sqrt{  \alpha L \lambda_p (l_c^2+4 \omega^2) }} + \frac{  \sqrt{\alpha L \lambda_p (l_c^2+4 \omega^2)}}{2 \omega l_c \sqrt{2 \pi}} \right]^2 
\end{equation}
with $L \approx 0.9mm$ (cristal thickness), $\lambda_p \approx 405$nm (pump wavelength), $\alpha = 0.455$~\cite{chan_transverse_2007} and $\omega \approx 89 \mu m$ (pump waist). Despite the many approximations that have been made and taking into accounts experimental uncertainties, we observe an excellent agreement between theoretically expected values of $K$ and those measured experimentally. Knowing the characteristics of the crystal and the pump therefore allows predicting reasonably well the degree of entanglement of the source. For a given crystal, we show that manipulating the pump coherence using rotating random diffusers enable the deterministic control of the degree of entanglement in the two-photon field generated. 

The future of quantum optical technologies depends on our capacity to detect~\cite{reichert_massively_2018,defienne_general_2018} and manipulate photons~\cite{defienne_adaptive_2018,peng_manipulation_2018}, but it also crucially relies on our ability to generate photons with properties adapted to specific application. In our work, we show how to produce spatially-entangled photons with specific degree of entanglement by controlling the spatial coherence of the pump beam with rotating random diffusers. For this, we investigated the fundamental transfer of coherence between the pump and the down-converted field and showed a good agreement with the theory~\cite{jha_spatial_2010,giese_influence_2018}. This novel source may play an important role in free-space quantum communications, since it has been recently shown in theory that a two-photon field is less susceptible to atmospheric turbulence when it was generated by a partial spatially coherent beam~\cite{qiu_influence_2012}. In this respect, the use of a spatial light modulator in place of the random diffusers will be the next natural step to enable tailoring entanglement in real-time and use it as a tunable parameter to produce quantum states that are optimal for a given protocol and strength of turbulence. Incoherent two-photon illumination could also plays an important role in optical imaging to improve resolution~\cite{hong_two-photon_2018}. Finally, this work may have technological impact as it paves the way towards the development of cheap and compact photon-pairs source using Light Emitting Diodes as pump beams~\cite{salter_entangled-light-emitting_2010}.

\bibliography{Biblio}

\clearpage

\appendix

\section{Theoretical model}

\subsection{Joint probability distribution in momentum-space $\Gamma(\vec{k_1},\vec{k_2})$}
As demonstrated in~\cite{giese_influence_2018} (Equation B.8), the joint probability distribution $\Gamma(\vec{k_1},\vec{k_2})$ for a partial spatially coherent pump beam is written as
\begin{equation}
\label{eqSM1}
\Gamma(\vec{k_1},\vec{k_2}) \sim |\tilde{\chi}(|\vec{k_1}-\vec{k_2}|^2)|^2 \tilde{V}(\vec{k_1}+\vec{k_2},\vec{k_1}+\vec{k_2})
\end{equation} 
where $\tilde{\chi}$ is the phase-matching function and $\tilde{V}$ is the transverse momentum-correlation function of the pump field. In our work, we use two distinct approximations:
\begin{itemize}
\item A Gaussian approximation~\citep{fedorov_gaussian_2009} for $\tilde{\chi}$:
\begin{equation}
\label{eqSM2}
|\tilde{\chi}(|\vec{k_1}-\vec{k_2}|^2)|^2 \sim \exp \left[-  \frac{\sigma_r^2 |\vec{k_1}-\vec{k_2}|^2}{2} \right]
\end{equation}
where $\sigma_r = \sqrt{\alpha L \lambda_p/(2 \pi)}$, with $\lambda_p$ is the pump wavelength, $L$ the crystal length and $\alpha = 0.455$~\cite{chan_transverse_2007}.
\item A Gaussian-Schell approximation~\cite{mandel_coherence_1965} for the partial spatially coherent pump beam, which results in $\tilde{V}$ being written as
\begin{equation}
\label{eqSM3}
\tilde{V}(\vec{k},\vec{k'}) \sim \exp \left[- \frac{\omega^2 |\vec{k}-\vec{k'}|^2}{2} - \frac{ |\vec{k}+\vec{k'}|^2}{8 \sigma_k^2} \right]
\end{equation}
where $\sigma_k = \sqrt{1/l_c^2+1/(4 \omega^2)}$, with $l_c$ the coherence length of the pump and $\omega$ its waist.
\end{itemize}
Combining Equations~\ref{eqSM2}, \ref{eqSM3} and~\ref{eqSM1} leads to Equation~\ref{eq1}. 

\subsection{Sum-coordinate projection of $\Gamma(\vec{k_1},\vec{k_2})$}
The sum-coordinate projection of $\Gamma$, denoted $P_+^{\Gamma}$, is calculated by integrating equation~\ref{eq1} along $\vec{k_1}+\vec{k_2}$ and takes the simple form
\begin{equation}
\label{eqSM31}
P_+^{\Gamma}(\vec{k_1}+\vec{k_2}) \sim \exp \left(- \frac{  |\vec{k_1}+\vec{k_2}|^2}{2 \sigma_k^2} \right)
\end{equation}
This model is used to fit the experimental data shown in Figure~\ref{Figure3}.b and to determine values of $\sigma_k$ reported in Table~\ref{table1}.

\section{Correlation-positions and $\sigma_r$ measurements}

\begin{figure}
\includegraphics[width=1 \columnwidth]{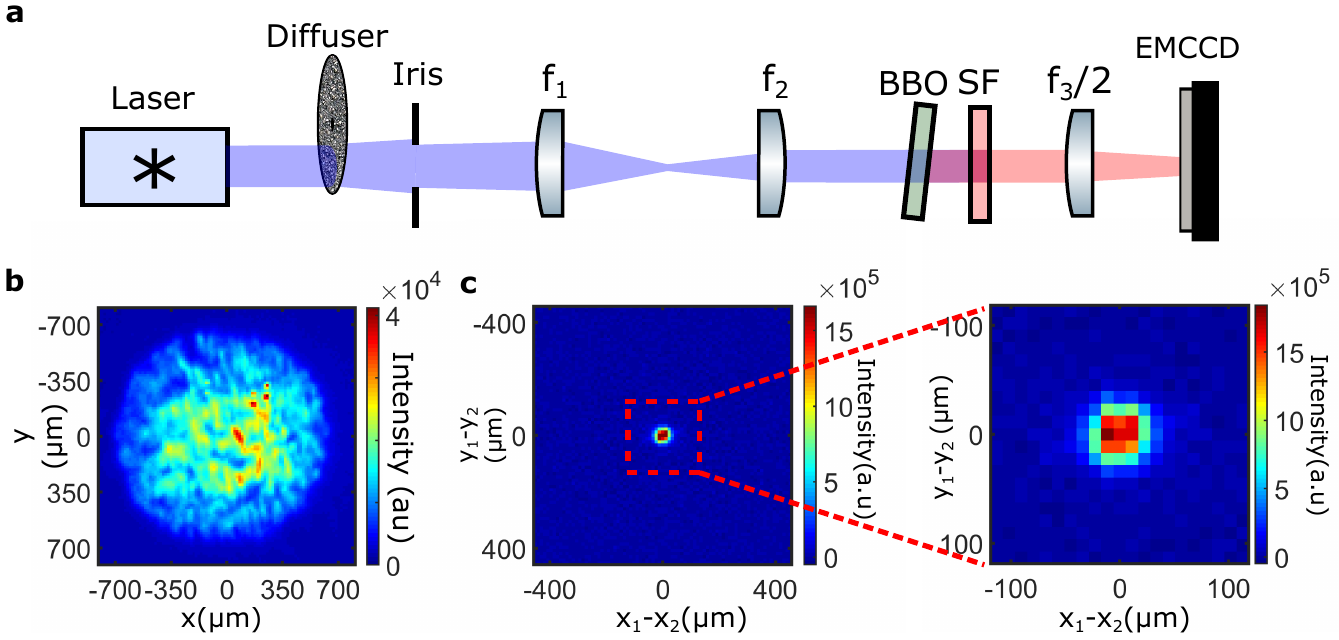}
\caption{\label{FigureSM1}   \textbf{(a)} Light emitted by a diode laser ($\lambda_p=405nm$) is scattered by a static thin diffuser (plastic sleeve) and illuminates a non-linear crystal of $\beta$-Baryum Borate (BBO) to produce spatially-entangled pairs of photons by type I SPDC. Spectral filters at $810 \pm 10$nm select near-degenerate photons. Lenses $f_1 =150$mm and $f_2 = 200$mm image an iris onto the cristal surface. When the diffuser is maintained fixed, the crystal is thus illuminated by a static speckle pattern.  White scale bar corresponds to $ 700 \mu m$. Positions of photons at the output surface of the crystal are imaged onto an EMCCD camera via a single-lens imaging system $f_3/2 = 20$mm. The direct intensity image \textbf{(b)} recorded by accumulating photons onto an EMCCD camera sensor is a speckle pattern. Minus-coordinate projection of the joint probability distribution  of photon-pairs \textbf{(c)} shows a strong peak at its center that reveals the strong correlations between positions of the pairs.}
\end{figure}

\subsection{Position-correlations}

Position-correlations between pairs of photons are observed by imaging the output surface of the crystal and measuring the joint probability distribution $\Gamma$, as shown in Figure~\ref{FigureSM1}.a. The diffuser is maintained static and is the same than the one used in Figure~\ref{Figure1}. The direct intensity image (Figure~\ref{FigureSM1}.b) is acquired by photons accumulation on the camera sensor and shows an speckle structure. When measuring the joint probability ditribution $\Gamma$ with the EMCCD camera~\cite{reichert_massively_2018,defienne_general_2018}, its projection along the minus-coordinate diagonal shows a central peak (Figure~\ref{FigureSM1}.c). The minus-coordinate projection image represents the probablity of detecting two photons from a pair separated by a (oriented) distance $r_1-r_2$~\cite{moreau_realization_2012,tasca_imaging_2012}. The strong peak at the center is a clear signature of the strong correlations in position between pairs of photons. 

\subsection{$\sigma_r$ measurements using partially coherent pump beams}

\begin{figure*}
\includegraphics[width=1 \textwidth]{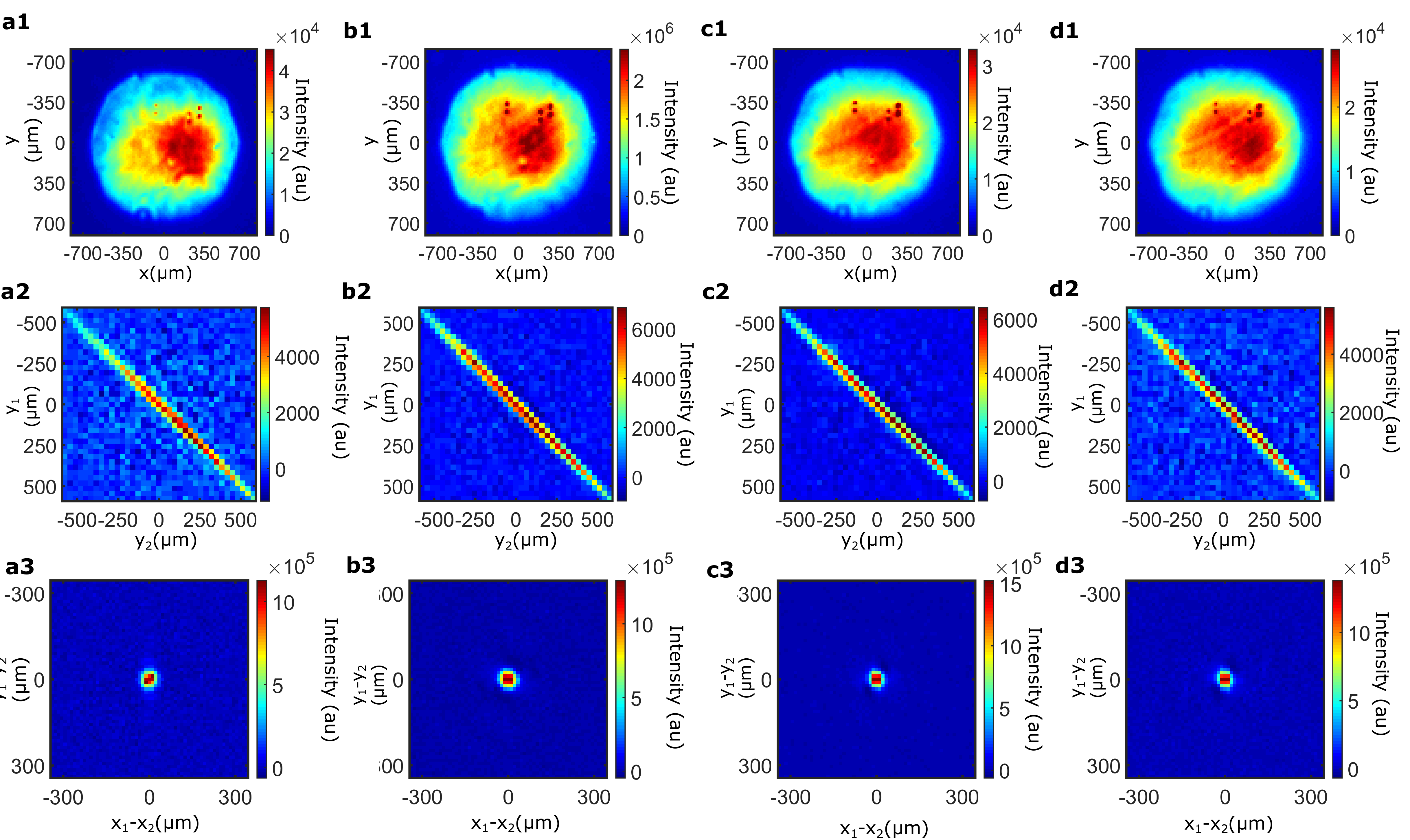}
\caption{\label{FigureSM2} Direct intensity of the down-converted field at the crystal plane is imaged onto the EMCCD camera using the experimental configuration described in Figure~\ref{FigureSM1}.a without diffuser \textbf{(a1)}, with a rotating diffuser composed by one layer of plastic sleeve \textbf{(b1)}, two layers \textbf{(c1)} and three layers \textbf{(d1)}. All intensity patterns are homogeneous and identical. When the camera is used to measure the joint probability distribution $\Gamma$, the $X_-$-projection of $\Gamma$ (see section 4) camera shows a very strong diagonal in all four cases: without diffuser \textbf{(a2)}, with one layer of plastic sleeve \textbf{(b2)}, two layers \textbf{(c2)} and three layers \textbf{(d2)}. These projections show that position-correlations do not depend on the coherence properties of the pump. Minus-coordinate projections of $\Gamma$ taken without diffuser \textbf{(a3)}, with one layer \textbf{(b3)}, two layers \textbf{(c3)} and three layers \textbf{(d3)} show the same peak at their center, which highlights the strong correlations between positions of the pairs. Fitting these images with the model of equation~\ref{eqSM41} allows detrermining values of $\sigma_r$ reported in Table~\ref{table1}. }
\end{figure*}

Values of $\sigma_r$ are determined using the experimental setup described Figure~\ref{FigureSM1}.a.  The same rotating diffusers (respectively composed by one, two and three layers of plastic sleeve) than those of Figure~\ref{Figure2} and Figure~\ref{Figure3} are used to generate partially coherent pump beams with different correlation lengths. Interestingly, Figure~\ref{FigureSM2} shows that neither the direct intensity images (Figure~\ref{FigureSM2}.a1-d1) nor the $X_-$-coordinate projections (Figure~\ref{FigureSM2}.a2-d2) depend on the coherence properties of the pump beam. The $X_-$-coordinate image represents the joint probability of detecting one photon at position $y_1$  ($x_1$ can take any possible values) and its twin with momentum $y_2$ and $x_2 \approx x_1$ (see section 4). The strong diagonal is a clear signature of position-correlations: both photons are always produced at the same position in the crystal during the SPDC process, and this property does not depend on the coherence properties of the pump beam.

Similarly to the calculations of section 1 and those of~\cite{giese_influence_2018}, the use of a Gaussian approximation~\cite{fedorov_gaussian_2009} and a Gaussian-Schell model~\cite{mandel_coherence_1965} allows writing the joint probability distribution $\Gamma(\vec{r_1},\vec{r_2})$ as
\begin{equation}
\label{eqSM4}
\Gamma(\vec{r_1},\vec{r_2}) \sim \exp \left(- \frac{  |\vec{r_1}-\vec{r_2}|^2}{2 \beta \sigma_r^2 } \right) \exp \left(- 2 \omega^2 |\vec{r_1}+\vec{r_2}|^2   \right)
\end{equation}
where $\omega$ is the pump beam waist and $\beta = (\alpha+\alpha^{-1})/\alpha$ ($\alpha = 0.455$~\cite{chan_transverse_2007}). The minus-coordinate projection of $\Gamma$, denoted $P_-^{\Gamma}$, is calculated by integrating equation~\ref{eqSM4} along $\vec{r_1}-\vec{r_2}$ and takes the simple form
\begin{equation}
\label{eqSM41}
P_-^{\Gamma}(\vec{r_1}-\vec{r_2}) \sim \exp \left(- \frac{  |\vec{r_1}-\vec{r_2}|^2}{2 \beta \sigma_r^2 } \right)
\end{equation}

The minus-coordinate projection images acquired for different correlation lengths are shown in Figure~\ref{FigureSM2}.a3-d3. Values of $\sigma_r$ are determined by fitting the minus-coordinate images by equation~\ref{eqSM41} and are reported in Table~\ref{table1}.

\section{Pump beam analysis and coherence length $\ell_c$ measurement}

Properties of the pump beam, namely its waist $\omega$ and correlation length $\ell_c$, are measured using the two experimental configurations described in Figure~\ref{FigureSM3}.a and b.

\subsection{Intensity distribution of the pump beam in the crystal plane}

The intensity distribution of the pump beam in the crystal plane is measured using the experimental configuration described in Figure~\ref{FigureSM3}.b. Figures~\ref{FigureSM3}.c-f show results of four acquisitions performed without diffuser (Figure~\ref{FigureSM3}.c), with a rotating diffuser composed by one layer of plastic sleeve (Figure~\ref{FigureSM3}.d), two layers (Figure~\ref{FigureSM3}.e) and three layers (Figure~\ref{FigureSM3}.f). Since the diffusers rotate with a period much shorter than the acquisition time of the camera, the distribution of pump intensity at the crystal plane is homogenous and does not depend on the coherence properties of the pump.

\subsection{Beam waist and correlation length measurements}

\begin{figure*}
\includegraphics[width=1 \textwidth]{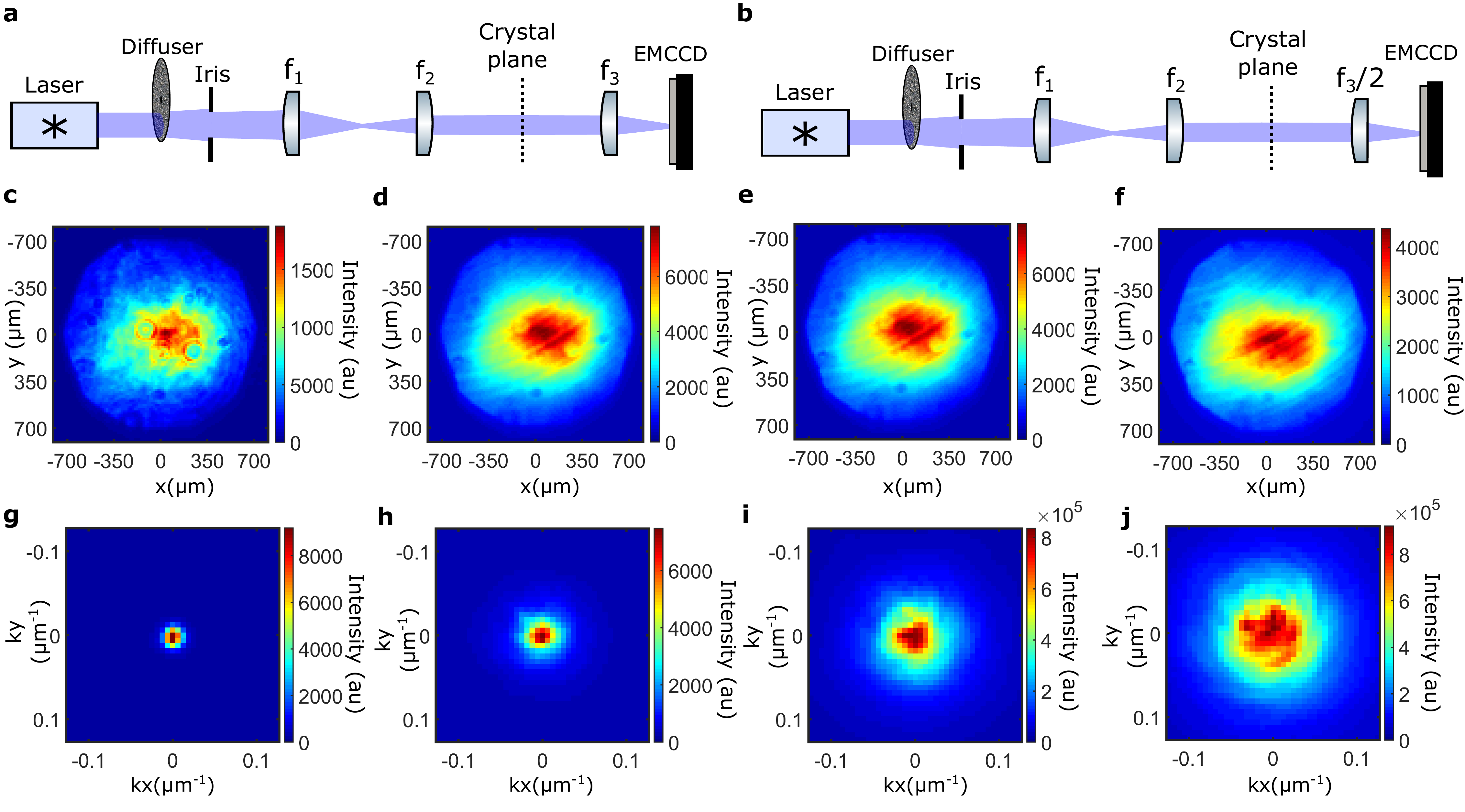}
\caption{\label{FigureSM3} \textbf{(a)} Apparatus used to Fourier-image the pump field at the crystal plane onto the camera. It is similar to the one shown in Figure~\ref{Figure1}.a without the crystal and all the filters. \textbf{(b)} Apparatus used to image the pump field at the crystal plane onto the camera. It is similar to the one shown in Figure~\ref{FigureSM1}.a without the crystal and all the filters. Using configuration \textbf{(b)}, intensity distribution of the pump beam at the crystal plane is imaged on the camera without diffuser \textbf{(c)}, with a rotating diffuser composed by one layer of plastic sleeve \textbf{(d)}, two layers \textbf{(e)} and three layers \textbf{(f)}. All intensity patterns are homogeneous and identical. Using configuration \textbf{(b)}, intensity distribution of the pump beam in the momentum space is imaged on the camera without diffuser \textbf{(g)}, with a rotating diffuser composed by one layer of plastic sleeve \textbf{(h)}, two layers \textbf{(i)} and three layers \textbf{(j)}. Without diffuser, the pump beam is focused onto the camera and the width of the peak $\sigma_{p_0}$ is used to estimate the beam waist $\omega = 1/\sigma_{p_0} \approx 89 \mu$m. When rotating diffusers are inserted, the peak gets broaden and its width $\sigma_p$ provides an estimation of the correlation length $\ell_c$ using the formula $\ell_c = 2 / \sqrt{\sigma_p^2-\sigma_{p_0}^2}$. Values of $\ell_c$ are reported in Table~\ref{table1}. }  
\end{figure*}

Measurements of $\omega$ and $\ell_c$ are performed using the experimental configuration of Figure~\ref{Figure1}.a. In this case, the pump field at the crystal plane is Fourier-imaged onto the EMCCD camera via lens $f_3$. Figures~\ref{FigureSM3}.g-j show four direct intensity images acquired respectively without diffuser (Figure~\ref{FigureSM3}.g), with a rotating diffuser composed by one layer of plastic sleeve (Figure~\ref{FigureSM3}.h), two layers (Figure~\ref{FigureSM3}.i) and three layers (Figure~\ref{FigureSM3}.j). For a perfectly coherent pump, the width of the focus (denoted $\sigma_{p_0}$) in Figure~\ref{FigureSM3}.g is inversely proportional to the beam waist $\omega$ 
\begin{equation}
\label{eqSM4}
\omega = \frac{1}{\sigma_{p_0}}
\end{equation}
Fitting this intensity distribution by Gaussian model provides an estimation of $\omega \approx 89 \mu$m. For partially coherent pump beams, intensity distributions in the Fourier domain shown in Figure~\ref{FigureSM3}.h-j are written as
\begin{equation}
\label{eqSM6}
I_p(\vec{k_p}) \sim \exp \left[ - \frac{|\vec{k_p}|^2}{2 \sigma_p^2} \right]
\end{equation}
where $\sigma_p = 2 \sqrt{1/\ell_c^2+1/(4 \omega^2)}$ (Gaussian-Schell model~\cite{mandel_coherence_1965}). Fitting these distributions with equation~\ref{eqSM6} allows determining $\sigma_p$ in each case and calculating $\ell_c$ with the formula
\begin{equation}
\label{eqSM4}
\ell_c = \frac{2}{\sqrt{\sigma_p^2-\sigma_{p_0}^2}}
\end{equation}
Values of $\ell_c$ are reported in Table~\ref{table1}.
 
\section{Image processing}

\subsection{Measurement process} We use an EMCCD Andor Ixon Ultra 897 to measure the joint probability distribution $\Gamma$ of spatially entangled photon pairs using a technique described in~\cite{defienne_general_2018}. The camera was operated at $-60^{\circ}$C, with a horizontal pixel shift readout rate of $17$Mhz, a vertical pixel shift every $0.3\,\mu$s and a vertical clock amplitude voltage of $+4$V above the factory setting. When the camera is illuminated by photon pairs, a large set of images is first collected using an exposure time chosen to have an intensity per pixel approximately $5$ times larger than mean value of the noise ($\sim 171$ grey values). No threshold is applied. Processing the set of images using the fomula provided in~\cite{defienne_general_2018} finally enables to reconstruct $\Gamma$. 

\subsection{Projections of the joint probability distribution} In our experiment, $\Gamma$ takes the form of a 4-dimensional matrix containing $(75 \times 75)^2 \sim 10^{8}$ elements, where $75 \times 75$ corresponds the size of the illuminated region of the camera sensor. The information content of $\Gamma$ is analyzed using four types of projections:

\begin{enumerate}
\item The sum-coordinate projection, defined as
\begin{equation}
P_+^{\Gamma}(\boldsymbol{\vec{k}_{+}}) = \sum_{\vec{k}} \Gamma(\vec{k}_{+}-\vec{k},\vec{k})
\end{equation}
It represents the probability of detecting pairs of photons generated in all symmetric directions relative to the mean momentum $\vec{k}_{+}$.
\item The minus-coordinate projection, defined as
\begin{equation}
P_-^{\Gamma}(\boldsymbol{\vec{r}_{-}}) = \sum_{\vec{r}} \Gamma(\vec{r}_{-}+\vec{r},\vec{r})
\end{equation}
It represents the probability for two photons of a pair to be detected in coincidence between pairs of pixels separated by an oriented distance ${\vec{r}_-}$.
\item The $X_+$-coordinate projection, defined as
\begin{align}
P_{X+}^{\Gamma} (k_{y_1},k_{y_2}) &= \sum_{k_{x}} \Gamma(k_{y_1},k_{y_2}|k_{x},-k_{x}) \\
&= \sum_{k_{x}} \frac{\Gamma(k_{y_1},k_{y_2},k_{x},-k_{x})}{\sum_{k_{x_1},k_{x_2}} \Gamma(k_{y_1},k_{y_2},k_{x_1},k_{x_2})}
\end{align}

It represents the probability of detecting one photon with momentum $k_{y_1}$ (with no constraints on $k_{x_1}$) given that the other is detected with a momentum $k_{y_2}$ and $k_{x_2}=-k_{x_1}$ [symetric columns].
\item The $X_-$-coordinate projection, defined as
\begin{align}
P_{X-}^{\Gamma} ({y_1},{y_2}) &= \sum_{{x}} \Gamma(y_1,y_2|x,x+1) \\
&= \sum_{{x}} \frac{\Gamma({y_1},{y_2},{x},{x}+1)}{\sum_{{x_1},{x_2}} \Gamma({y_1},{y_2},{x_1},{x_2})}
\end{align}
It represents the probability of detecting one photon at position ${y_1}$ (with no constraints on ${x_1}$) given that the other is detected with a momentum ${y_2}$ and ${x_2}={x_1}+1$ [adjacent columns].
\end{enumerate}

\bibliography{Biblio}

\begin{thebibliography}{40}%
\makeatletter
\providecommand \@ifxundefined [1]{%
 \@ifx{#1\undefined}
}%
\providecommand \@ifnum [1]{%
 \ifnum #1\expandafter \@firstoftwo
 \else \expandafter \@secondoftwo
 \fi
}%
\providecommand \@ifx [1]{%
 \ifx #1\expandafter \@firstoftwo
 \else \expandafter \@secondoftwo
 \fi
}%
\providecommand \natexlab [1]{#1}%
\providecommand \enquote  [1]{``#1''}%
\providecommand \bibnamefont  [1]{#1}%
\providecommand \bibfnamefont [1]{#1}%
\providecommand \citenamefont [1]{#1}%
\providecommand \href@noop [0]{\@secondoftwo}%
\providecommand \href [0]{\begingroup \@sanitize@url \@href}%
\providecommand \@href[1]{\@@startlink{#1}\@@href}%
\providecommand \@@href[1]{\endgroup#1\@@endlink}%
\providecommand \@sanitize@url [0]{\catcode `\\12\catcode `\$12\catcode
  `\&12\catcode `\#12\catcode `\^12\catcode `\_12\catcode `\%12\relax}%
\providecommand \@@startlink[1]{}%
\providecommand \@@endlink[0]{}%
\providecommand \url  [0]{\begingroup\@sanitize@url \@url }%
\providecommand \@url [1]{\endgroup\@href {#1}{\urlprefix }}%
\providecommand \urlprefix  [0]{URL }%
\providecommand \Eprint [0]{\href }%
\providecommand \doibase [0]{http://dx.doi.org/}%
\providecommand \selectlanguage [0]{\@gobble}%
\providecommand \bibinfo  [0]{\@secondoftwo}%
\providecommand \bibfield  [0]{\@secondoftwo}%
\providecommand \translation [1]{[#1]}%
\providecommand \BibitemOpen [0]{}%
\providecommand \bibitemStop [0]{}%
\providecommand \bibitemNoStop [0]{.\EOS\space}%
\providecommand \EOS [0]{\spacefactor3000\relax}%
\providecommand \BibitemShut  [1]{\csname bibitem#1\endcsname}%
\let\auto@bib@innerbib\@empty
\bibitem [{\citenamefont {Brendel}\ \emph {et~al.}(1999)\citenamefont
  {Brendel}, \citenamefont {Gisin}, \citenamefont {Tittel},\ and\ \citenamefont
  {Zbinden}}]{brendel_pulsed_1999}%
  \BibitemOpen
  \bibfield  {author} {\bibinfo {author} {\bibfnamefont {J.}~\bibnamefont
  {Brendel}}, \bibinfo {author} {\bibfnamefont {N.}~\bibnamefont {Gisin}},
  \bibinfo {author} {\bibfnamefont {W.}~\bibnamefont {Tittel}}, \ and\ \bibinfo
  {author} {\bibfnamefont {H.}~\bibnamefont {Zbinden}},\ }\href {\doibase
  10.1103/PhysRevLett.82.2594} {\bibfield  {journal} {\bibinfo  {journal}
  {Physical Review Letters}\ }\textbf {\bibinfo {volume} {82}},\ \bibinfo
  {pages} {2594} (\bibinfo {year} {1999})}\BibitemShut {NoStop}%
\bibitem [{\citenamefont {Kwiat}\ \emph {et~al.}(1995)\citenamefont {Kwiat},
  \citenamefont {Mattle}, \citenamefont {Weinfurter}, \citenamefont
  {Zeilinger}, \citenamefont {Sergienko},\ and\ \citenamefont
  {Shih}}]{kwiat_new_1995}%
  \BibitemOpen
  \bibfield  {author} {\bibinfo {author} {\bibfnamefont {P.~G.}\ \bibnamefont
  {Kwiat}}, \bibinfo {author} {\bibfnamefont {K.}~\bibnamefont {Mattle}},
  \bibinfo {author} {\bibfnamefont {H.}~\bibnamefont {Weinfurter}}, \bibinfo
  {author} {\bibfnamefont {A.}~\bibnamefont {Zeilinger}}, \bibinfo {author}
  {\bibfnamefont {A.~V.}\ \bibnamefont {Sergienko}}, \ and\ \bibinfo {author}
  {\bibfnamefont {Y.}~\bibnamefont {Shih}},\ }\href {\doibase
  10.1103/PhysRevLett.75.4337} {\bibfield  {journal} {\bibinfo  {journal}
  {Physical Review Letters}\ }\textbf {\bibinfo {volume} {75}},\ \bibinfo
  {pages} {4337} (\bibinfo {year} {1995})}\BibitemShut {NoStop}%
\bibitem [{\citenamefont {Howell}\ \emph {et~al.}(2004)\citenamefont {Howell},
  \citenamefont {Bennink}, \citenamefont {Bentley},\ and\ \citenamefont
  {Boyd}}]{howell_realization_2004}%
  \BibitemOpen
  \bibfield  {author} {\bibinfo {author} {\bibfnamefont {J.~C.}\ \bibnamefont
  {Howell}}, \bibinfo {author} {\bibfnamefont {R.~S.}\ \bibnamefont {Bennink}},
  \bibinfo {author} {\bibfnamefont {S.~J.}\ \bibnamefont {Bentley}}, \ and\
  \bibinfo {author} {\bibfnamefont {R.~W.}\ \bibnamefont {Boyd}},\ }\href
  {\doibase 10.1103/PhysRevLett.92.210403} {\bibfield  {journal} {\bibinfo
  {journal} {Physical Review Letters}\ }\textbf {\bibinfo {volume} {92}},\
  \bibinfo {pages} {210403} (\bibinfo {year} {2004})}\BibitemShut {NoStop}%
\bibitem [{\citenamefont {Aspect}\ \emph {et~al.}(1982)\citenamefont {Aspect},
  \citenamefont {Grangier},\ and\ \citenamefont
  {Roger}}]{aspect_experimental_1982}%
  \BibitemOpen
  \bibfield  {author} {\bibinfo {author} {\bibfnamefont {A.}~\bibnamefont
  {Aspect}}, \bibinfo {author} {\bibfnamefont {P.}~\bibnamefont {Grangier}}, \
  and\ \bibinfo {author} {\bibfnamefont {G.}~\bibnamefont {Roger}},\ }\href
  {http://journals.aps.org/prl/abstract/10.1103/PhysRevLett.49.91} {\bibfield
  {journal} {\bibinfo  {journal} {Physical review letters}\ }\textbf {\bibinfo
  {volume} {49}},\ \bibinfo {pages} {91} (\bibinfo {year} {1982})}\BibitemShut
  {NoStop}%
\bibitem [{\citenamefont {Liao}\ \emph {et~al.}(2017)\citenamefont {Liao},
  \citenamefont {Cai}, \citenamefont {Liu}, \citenamefont {Zhang},
  \citenamefont {Li}, \citenamefont {Ren}, \citenamefont {Yin}, \citenamefont
  {Shen}, \citenamefont {Cao}, \citenamefont {Li}, \citenamefont {Li},
  \citenamefont {Chen}, \citenamefont {Sun}, \citenamefont {Jia}, \citenamefont
  {Wu}, \citenamefont {Jiang}, \citenamefont {Wang}, \citenamefont {Huang},
  \citenamefont {Wang}, \citenamefont {Zhou}, \citenamefont {Deng},
  \citenamefont {Xi}, \citenamefont {Ma}, \citenamefont {Hu}, \citenamefont
  {Zhang}, \citenamefont {Chen}, \citenamefont {Liu}, \citenamefont {Wang},
  \citenamefont {Zhu}, \citenamefont {Lu}, \citenamefont {Shu}, \citenamefont
  {Peng}, \citenamefont {Wang},\ and\ \citenamefont
  {Pan}}]{liao_satellite--ground_2017}%
  \BibitemOpen
  \bibfield  {author} {\bibinfo {author} {\bibfnamefont {S.-K.}\ \bibnamefont
  {Liao}}, \bibinfo {author} {\bibfnamefont {W.-Q.}\ \bibnamefont {Cai}},
  \bibinfo {author} {\bibfnamefont {W.-Y.}\ \bibnamefont {Liu}}, \bibinfo
  {author} {\bibfnamefont {L.}~\bibnamefont {Zhang}}, \bibinfo {author}
  {\bibfnamefont {Y.}~\bibnamefont {Li}}, \bibinfo {author} {\bibfnamefont
  {J.-G.}\ \bibnamefont {Ren}}, \bibinfo {author} {\bibfnamefont
  {J.}~\bibnamefont {Yin}}, \bibinfo {author} {\bibfnamefont {Q.}~\bibnamefont
  {Shen}}, \bibinfo {author} {\bibfnamefont {Y.}~\bibnamefont {Cao}}, \bibinfo
  {author} {\bibfnamefont {Z.-P.}\ \bibnamefont {Li}}, \bibinfo {author}
  {\bibfnamefont {F.-Z.}\ \bibnamefont {Li}}, \bibinfo {author} {\bibfnamefont
  {X.-W.}\ \bibnamefont {Chen}}, \bibinfo {author} {\bibfnamefont {L.-H.}\
  \bibnamefont {Sun}}, \bibinfo {author} {\bibfnamefont {J.-J.}\ \bibnamefont
  {Jia}}, \bibinfo {author} {\bibfnamefont {J.-C.}\ \bibnamefont {Wu}},
  \bibinfo {author} {\bibfnamefont {X.-J.}\ \bibnamefont {Jiang}}, \bibinfo
  {author} {\bibfnamefont {J.-F.}\ \bibnamefont {Wang}}, \bibinfo {author}
  {\bibfnamefont {Y.-M.}\ \bibnamefont {Huang}}, \bibinfo {author}
  {\bibfnamefont {Q.}~\bibnamefont {Wang}}, \bibinfo {author} {\bibfnamefont
  {Y.-L.}\ \bibnamefont {Zhou}}, \bibinfo {author} {\bibfnamefont
  {L.}~\bibnamefont {Deng}}, \bibinfo {author} {\bibfnamefont {T.}~\bibnamefont
  {Xi}}, \bibinfo {author} {\bibfnamefont {L.}~\bibnamefont {Ma}}, \bibinfo
  {author} {\bibfnamefont {T.}~\bibnamefont {Hu}}, \bibinfo {author}
  {\bibfnamefont {Q.}~\bibnamefont {Zhang}}, \bibinfo {author} {\bibfnamefont
  {Y.-A.}\ \bibnamefont {Chen}}, \bibinfo {author} {\bibfnamefont {N.-L.}\
  \bibnamefont {Liu}}, \bibinfo {author} {\bibfnamefont {X.-B.}\ \bibnamefont
  {Wang}}, \bibinfo {author} {\bibfnamefont {Z.-C.}\ \bibnamefont {Zhu}},
  \bibinfo {author} {\bibfnamefont {C.-Y.}\ \bibnamefont {Lu}}, \bibinfo
  {author} {\bibfnamefont {R.}~\bibnamefont {Shu}}, \bibinfo {author}
  {\bibfnamefont {C.-Z.}\ \bibnamefont {Peng}}, \bibinfo {author}
  {\bibfnamefont {J.-Y.}\ \bibnamefont {Wang}}, \ and\ \bibinfo {author}
  {\bibfnamefont {J.-W.}\ \bibnamefont {Pan}},\ }\href {\doibase
  10.1038/nature23655} {\bibfield  {journal} {\bibinfo  {journal} {Nature}\
  }\textbf {\bibinfo {volume} {549}},\ \bibinfo {pages} {43} (\bibinfo {year}
  {2017})}\BibitemShut {NoStop}%
\bibitem [{\citenamefont {Walborn}\ \emph {et~al.}(2010)\citenamefont
  {Walborn}, \citenamefont {Monken}, \citenamefont {Pádua},\ and\
  \citenamefont {Souto~Ribeiro}}]{walborn_spatial_2010}%
  \BibitemOpen
  \bibfield  {author} {\bibinfo {author} {\bibfnamefont {S.~P.}\ \bibnamefont
  {Walborn}}, \bibinfo {author} {\bibfnamefont {C.~H.}\ \bibnamefont {Monken}},
  \bibinfo {author} {\bibfnamefont {S.}~\bibnamefont {Pádua}}, \ and\ \bibinfo
  {author} {\bibfnamefont {P.~H.}\ \bibnamefont {Souto~Ribeiro}},\ }\href
  {\doibase 10.1016/j.physrep.2010.06.003} {\bibfield  {journal} {\bibinfo
  {journal} {Physics Reports}\ }\textbf {\bibinfo {volume} {495}},\ \bibinfo
  {pages} {87} (\bibinfo {year} {2010})}\BibitemShut {NoStop}%
\bibitem [{\citenamefont {Tasca}\ \emph {et~al.}(2011)\citenamefont {Tasca},
  \citenamefont {Gomes}, \citenamefont {Toscano}, \citenamefont
  {Souto~Ribeiro},\ and\ \citenamefont
  {Walborn}}]{tasca_continuous-variable_2011}%
  \BibitemOpen
  \bibfield  {author} {\bibinfo {author} {\bibfnamefont {D.~S.}\ \bibnamefont
  {Tasca}}, \bibinfo {author} {\bibfnamefont {R.~M.}\ \bibnamefont {Gomes}},
  \bibinfo {author} {\bibfnamefont {F.}~\bibnamefont {Toscano}}, \bibinfo
  {author} {\bibfnamefont {P.~H.}\ \bibnamefont {Souto~Ribeiro}}, \ and\
  \bibinfo {author} {\bibfnamefont {S.~P.}\ \bibnamefont {Walborn}},\ }\href
  {\doibase 10.1103/PhysRevA.83.052325} {\bibfield  {journal} {\bibinfo
  {journal} {Physical Review A}\ }\textbf {\bibinfo {volume} {83}},\ \bibinfo
  {pages} {052325} (\bibinfo {year} {2011})}\BibitemShut {NoStop}%
\bibitem [{\citenamefont {Walborn}\ \emph {et~al.}(2006)\citenamefont
  {Walborn}, \citenamefont {Lemelle}, \citenamefont {Almeida},\ and\
  \citenamefont {Ribeiro}}]{walborn_quantum_2006}%
  \BibitemOpen
  \bibfield  {author} {\bibinfo {author} {\bibfnamefont {S.~P.}\ \bibnamefont
  {Walborn}}, \bibinfo {author} {\bibfnamefont {D.~S.}\ \bibnamefont
  {Lemelle}}, \bibinfo {author} {\bibfnamefont {M.~P.}\ \bibnamefont
  {Almeida}}, \ and\ \bibinfo {author} {\bibfnamefont {P.~S.}\ \bibnamefont
  {Ribeiro}},\ }\href@noop {} {\bibfield  {journal} {\bibinfo  {journal}
  {Physical review letters}\ }\textbf {\bibinfo {volume} {96}},\ \bibinfo
  {pages} {090501} (\bibinfo {year} {2006})}\BibitemShut {NoStop}%
\bibitem [{\citenamefont {Pittman}\ \emph {et~al.}(1995)\citenamefont
  {Pittman}, \citenamefont {Shih}, \citenamefont {Strekalov},\ and\
  \citenamefont {Sergienko}}]{pittman_optical_1995}%
  \BibitemOpen
  \bibfield  {author} {\bibinfo {author} {\bibfnamefont {T.~B.}\ \bibnamefont
  {Pittman}}, \bibinfo {author} {\bibfnamefont {Y.~H.}\ \bibnamefont {Shih}},
  \bibinfo {author} {\bibfnamefont {D.~V.}\ \bibnamefont {Strekalov}}, \ and\
  \bibinfo {author} {\bibfnamefont {A.~V.}\ \bibnamefont {Sergienko}},\ }\href
  {\doibase 10.1103/PhysRevA.52.R3429} {\bibfield  {journal} {\bibinfo
  {journal} {Physical Review A}\ }\textbf {\bibinfo {volume} {52}},\ \bibinfo
  {pages} {R3429} (\bibinfo {year} {1995})}\BibitemShut {NoStop}%
\bibitem [{\citenamefont {Brida}\ \emph {et~al.}(2010)\citenamefont {Brida},
  \citenamefont {Genovese},\ and\ \citenamefont
  {Berchera}}]{brida_experimental_2010}%
  \BibitemOpen
  \bibfield  {author} {\bibinfo {author} {\bibfnamefont {G.}~\bibnamefont
  {Brida}}, \bibinfo {author} {\bibfnamefont {M.}~\bibnamefont {Genovese}}, \
  and\ \bibinfo {author} {\bibfnamefont {I.~R.}\ \bibnamefont {Berchera}},\
  }\href {\doibase 10.1038/nphoton.2010.29} {\bibfield  {journal} {\bibinfo
  {journal} {Nature Photonics}\ }\textbf {\bibinfo {volume} {4}},\ \bibinfo
  {pages} {227} (\bibinfo {year} {2010})}\BibitemShut {NoStop}%
\bibitem [{\citenamefont {Xu}\ \emph {et~al.}(2015)\citenamefont {Xu},
  \citenamefont {Song}, \citenamefont {Li}, \citenamefont {Zhang},
  \citenamefont {Wang}, \citenamefont {Xiong},\ and\ \citenamefont
  {Wang}}]{xu_experimental_2015}%
  \BibitemOpen
  \bibfield  {author} {\bibinfo {author} {\bibfnamefont {D.-Q.}\ \bibnamefont
  {Xu}}, \bibinfo {author} {\bibfnamefont {X.-B.}\ \bibnamefont {Song}},
  \bibinfo {author} {\bibfnamefont {H.-G.}\ \bibnamefont {Li}}, \bibinfo
  {author} {\bibfnamefont {D.-J.}\ \bibnamefont {Zhang}}, \bibinfo {author}
  {\bibfnamefont {H.-B.}\ \bibnamefont {Wang}}, \bibinfo {author}
  {\bibfnamefont {J.}~\bibnamefont {Xiong}}, \ and\ \bibinfo {author}
  {\bibfnamefont {K.}~\bibnamefont {Wang}},\ }\href {\doibase
  10.1063/1.4919131} {\bibfield  {journal} {\bibinfo  {journal} {Applied
  Physics Letters}\ }\textbf {\bibinfo {volume} {106}},\ \bibinfo {pages}
  {171104} (\bibinfo {year} {2015})}\BibitemShut {NoStop}%
\bibitem [{\citenamefont {Dixon}\ \emph {et~al.}(2012)\citenamefont {Dixon},
  \citenamefont {Howland}, \citenamefont {Schneeloch},\ and\ \citenamefont
  {Howell}}]{dixon_quantum_2012}%
  \BibitemOpen
  \bibfield  {author} {\bibinfo {author} {\bibfnamefont {P.~B.}\ \bibnamefont
  {Dixon}}, \bibinfo {author} {\bibfnamefont {G.~A.}\ \bibnamefont {Howland}},
  \bibinfo {author} {\bibfnamefont {J.}~\bibnamefont {Schneeloch}}, \ and\
  \bibinfo {author} {\bibfnamefont {J.~C.}\ \bibnamefont {Howell}},\ }\href
  {\doibase 10.1103/PhysRevLett.108.143603} {\bibfield  {journal} {\bibinfo
  {journal} {Physical Review Letters}\ }\textbf {\bibinfo {volume} {108}},\
  \bibinfo {pages} {143603} (\bibinfo {year} {2012})}\BibitemShut {NoStop}%
\bibitem [{\citenamefont {Reichert}\ \emph {et~al.}(2017)\citenamefont
  {Reichert}, \citenamefont {Defienne}, \citenamefont {Sun},\ and\
  \citenamefont {Fleischer}}]{reichert_biphoton_2017}%
  \BibitemOpen
  \bibfield  {author} {\bibinfo {author} {\bibfnamefont {M.}~\bibnamefont
  {Reichert}}, \bibinfo {author} {\bibfnamefont {H.}~\bibnamefont {Defienne}},
  \bibinfo {author} {\bibfnamefont {X.}~\bibnamefont {Sun}}, \ and\ \bibinfo
  {author} {\bibfnamefont {J.~W.}\ \bibnamefont {Fleischer}},\ }\href {\doibase
  10.1088/2040-8986/aa6175} {\bibfield  {journal} {\bibinfo  {journal} {Journal
  of Optics}\ }\textbf {\bibinfo {volume} {19}},\ \bibinfo {pages} {044004}
  (\bibinfo {year} {2017})}\BibitemShut {NoStop}%
\bibitem [{\citenamefont {Hong}\ and\ \citenamefont
  {Mandel}(1985)}]{hong_theory_1985}%
  \BibitemOpen
  \bibfield  {author} {\bibinfo {author} {\bibfnamefont {C.~K.}\ \bibnamefont
  {Hong}}\ and\ \bibinfo {author} {\bibfnamefont {L.}~\bibnamefont {Mandel}},\
  }\href {\doibase 10.1103/PhysRevA.31.2409} {\bibfield  {journal} {\bibinfo
  {journal} {Physical Review A}\ }\textbf {\bibinfo {volume} {31}},\ \bibinfo
  {pages} {2409} (\bibinfo {year} {1985})}\BibitemShut {NoStop}%
\bibitem [{\citenamefont {Rubin}(1996)}]{rubin_transverse_1996}%
  \BibitemOpen
  \bibfield  {author} {\bibinfo {author} {\bibfnamefont {M.~H.}\ \bibnamefont
  {Rubin}},\ }\href {\doibase 10.1103/PhysRevA.54.5349} {\bibfield  {journal}
  {\bibinfo  {journal} {Physical Review A}\ }\textbf {\bibinfo {volume} {54}},\
  \bibinfo {pages} {5349} (\bibinfo {year} {1996})}\BibitemShut {NoStop}%
\bibitem [{\citenamefont {Souto~Ribeiro}(1997)}]{souto_ribeiro_partial_1997}%
  \BibitemOpen
  \bibfield  {author} {\bibinfo {author} {\bibfnamefont {P.~H.}\ \bibnamefont
  {Souto~Ribeiro}},\ }\href {\doibase 10.1103/PhysRevA.56.4111} {\bibfield
  {journal} {\bibinfo  {journal} {Physical Review A}\ }\textbf {\bibinfo
  {volume} {56}},\ \bibinfo {pages} {4111} (\bibinfo {year}
  {1997})}\BibitemShut {NoStop}%
\bibitem [{\citenamefont {Joobeur}\ \emph {et~al.}(1996)\citenamefont
  {Joobeur}, \citenamefont {Saleh}, \citenamefont {Larchuk},\ and\
  \citenamefont {Teich}}]{joobeur_coherence_1996}%
  \BibitemOpen
  \bibfield  {author} {\bibinfo {author} {\bibfnamefont {A.}~\bibnamefont
  {Joobeur}}, \bibinfo {author} {\bibfnamefont {B.~E.~A.}\ \bibnamefont
  {Saleh}}, \bibinfo {author} {\bibfnamefont {T.~S.}\ \bibnamefont {Larchuk}},
  \ and\ \bibinfo {author} {\bibfnamefont {M.~C.}\ \bibnamefont {Teich}},\
  }\href {\doibase 10.1103/PhysRevA.53.4360} {\bibfield  {journal} {\bibinfo
  {journal} {Physical Review A}\ }\textbf {\bibinfo {volume} {53}},\ \bibinfo
  {pages} {4360} (\bibinfo {year} {1996})}\BibitemShut {NoStop}%
\bibitem [{\citenamefont {Fonseca}\ \emph {et~al.}(1999)\citenamefont
  {Fonseca}, \citenamefont {Monken}, \citenamefont {Pádua},\ and\
  \citenamefont {Barbosa}}]{fonseca_transverse_1999}%
  \BibitemOpen
  \bibfield  {author} {\bibinfo {author} {\bibfnamefont {E.~J.~S.}\
  \bibnamefont {Fonseca}}, \bibinfo {author} {\bibfnamefont {C.~H.}\
  \bibnamefont {Monken}}, \bibinfo {author} {\bibfnamefont {S.}~\bibnamefont
  {Pádua}}, \ and\ \bibinfo {author} {\bibfnamefont {G.~A.}\ \bibnamefont
  {Barbosa}},\ }\href {\doibase 10.1103/PhysRevA.59.1608} {\bibfield  {journal}
  {\bibinfo  {journal} {Physical Review A}\ }\textbf {\bibinfo {volume} {59}},\
  \bibinfo {pages} {1608} (\bibinfo {year} {1999})}\BibitemShut {NoStop}%
\bibitem [{\citenamefont {Saleh}\ \emph {et~al.}(2005)\citenamefont {Saleh},
  \citenamefont {Teich},\ and\ \citenamefont {Sergienko}}]{saleh_wolf_2005}%
  \BibitemOpen
  \bibfield  {author} {\bibinfo {author} {\bibfnamefont {B.~E.~A.}\
  \bibnamefont {Saleh}}, \bibinfo {author} {\bibfnamefont {M.~C.}\ \bibnamefont
  {Teich}}, \ and\ \bibinfo {author} {\bibfnamefont {A.~V.}\ \bibnamefont
  {Sergienko}},\ }\href {\doibase 10.1103/PhysRevLett.94.223601} {\bibfield
  {journal} {\bibinfo  {journal} {Physical Review Letters}\ }\textbf {\bibinfo
  {volume} {94}},\ \bibinfo {pages} {223601} (\bibinfo {year}
  {2005})}\BibitemShut {NoStop}%
\bibitem [{\citenamefont {Monken}\ \emph {et~al.}(1998)\citenamefont {Monken},
  \citenamefont {Ribeiro},\ and\ \citenamefont
  {Pádua}}]{monken_transfer_1998}%
  \BibitemOpen
  \bibfield  {author} {\bibinfo {author} {\bibfnamefont {C.~H.}\ \bibnamefont
  {Monken}}, \bibinfo {author} {\bibfnamefont {P.~H.~S.}\ \bibnamefont
  {Ribeiro}}, \ and\ \bibinfo {author} {\bibfnamefont {S.}~\bibnamefont
  {Pádua}},\ }\href {\doibase 10.1103/PhysRevA.57.3123} {\bibfield  {journal}
  {\bibinfo  {journal} {Physical Review A}\ }\textbf {\bibinfo {volume} {57}},\
  \bibinfo {pages} {3123} (\bibinfo {year} {1998})}\BibitemShut {NoStop}%
\bibitem [{\citenamefont {Kulkarni}\ \emph {et~al.}(2017)\citenamefont
  {Kulkarni}, \citenamefont {Kumar},\ and\ \citenamefont
  {Jha}}]{kulkarni_transfer_2017}%
  \BibitemOpen
  \bibfield  {author} {\bibinfo {author} {\bibfnamefont {G.}~\bibnamefont
  {Kulkarni}}, \bibinfo {author} {\bibfnamefont {P.}~\bibnamefont {Kumar}}, \
  and\ \bibinfo {author} {\bibfnamefont {A.~K.}\ \bibnamefont {Jha}},\ }\href
  {\doibase 10.1364/JOSAB.34.001637} {\bibfield  {journal} {\bibinfo  {journal}
  {JOSA B}\ }\textbf {\bibinfo {volume} {34}},\ \bibinfo {pages} {1637}
  (\bibinfo {year} {2017})}\BibitemShut {NoStop}%
\bibitem [{\citenamefont {Ismail}\ \emph {et~al.}(2017)\citenamefont {Ismail},
  \citenamefont {Joshi},\ and\ \citenamefont
  {Petruccione}}]{ismail_polarization-entangled_2017}%
  \BibitemOpen
  \bibfield  {author} {\bibinfo {author} {\bibfnamefont {Y.}~\bibnamefont
  {Ismail}}, \bibinfo {author} {\bibfnamefont {S.}~\bibnamefont {Joshi}}, \
  and\ \bibinfo {author} {\bibfnamefont {F.}~\bibnamefont {Petruccione}},\
  }\href@noop {} {\bibfield  {journal} {\bibinfo  {journal} {Scientific
  reports}\ }\textbf {\bibinfo {volume} {7}},\ \bibinfo {pages} {12091}
  (\bibinfo {year} {2017})}\BibitemShut {NoStop}%
\bibitem [{\citenamefont {Jha}\ and\ \citenamefont
  {Boyd}(2010)}]{jha_spatial_2010}%
  \BibitemOpen
  \bibfield  {author} {\bibinfo {author} {\bibfnamefont {A.~K.}\ \bibnamefont
  {Jha}}\ and\ \bibinfo {author} {\bibfnamefont {R.~W.}\ \bibnamefont {Boyd}},\
  }\href {\doibase 10.1103/PhysRevA.81.013828} {\bibfield  {journal} {\bibinfo
  {journal} {Physical Review A}\ }\textbf {\bibinfo {volume} {81}},\ \bibinfo
  {pages} {013828} (\bibinfo {year} {2010})}\BibitemShut {NoStop}%
\bibitem [{\citenamefont {Olvera}\ and\ \citenamefont
  {Franke-Arnold}(2015)}]{olvera_two_2015}%
  \BibitemOpen
  \bibfield  {author} {\bibinfo {author} {\bibfnamefont {M.~A.}\ \bibnamefont
  {Olvera}}\ and\ \bibinfo {author} {\bibfnamefont {S.}~\bibnamefont
  {Franke-Arnold}},\ }\href@noop {} {\bibfield  {journal} {\bibinfo  {journal}
  {arXiv preprint arXiv:1507.08623}\ } (\bibinfo {year} {2015})}\BibitemShut
  {NoStop}%
\bibitem [{\citenamefont {Giese}\ \emph {et~al.}(2018)\citenamefont {Giese},
  \citenamefont {Fickler}, \citenamefont {Zhang}, \citenamefont {Chen},\ and\
  \citenamefont {Boyd}}]{giese_influence_2018}%
  \BibitemOpen
  \bibfield  {author} {\bibinfo {author} {\bibfnamefont {E.}~\bibnamefont
  {Giese}}, \bibinfo {author} {\bibfnamefont {R.}~\bibnamefont {Fickler}},
  \bibinfo {author} {\bibfnamefont {W.}~\bibnamefont {Zhang}}, \bibinfo
  {author} {\bibfnamefont {L.}~\bibnamefont {Chen}}, \ and\ \bibinfo {author}
  {\bibfnamefont {R.~W.}\ \bibnamefont {Boyd}},\ }\href {\doibase
  10.1088/1402-4896/aace12} {\bibfield  {journal} {\bibinfo  {journal} {Physica
  Scripta}\ }\textbf {\bibinfo {volume} {93}},\ \bibinfo {pages} {084001}
  (\bibinfo {year} {2018})}\BibitemShut {NoStop}%
\bibitem [{\citenamefont {Fedorov}\ \emph {et~al.}(2009)\citenamefont
  {Fedorov}, \citenamefont {Mikhailova},\ and\ \citenamefont
  {Volkov}}]{fedorov_gaussian_2009}%
  \BibitemOpen
  \bibfield  {author} {\bibinfo {author} {\bibfnamefont {M.~V.}\ \bibnamefont
  {Fedorov}}, \bibinfo {author} {\bibfnamefont {Y.~M.}\ \bibnamefont
  {Mikhailova}}, \ and\ \bibinfo {author} {\bibfnamefont {P.~A.}\ \bibnamefont
  {Volkov}},\ }\href@noop {} {\bibfield  {journal} {\bibinfo  {journal}
  {Journal of Physics B: Atomic, Molecular and Optical Physics}\ }\textbf
  {\bibinfo {volume} {42}},\ \bibinfo {pages} {175503} (\bibinfo {year}
  {2009})}\BibitemShut {NoStop}%
\bibitem [{\citenamefont {Reichert}\ \emph {et~al.}(2018)\citenamefont
  {Reichert}, \citenamefont {Defienne},\ and\ \citenamefont
  {Fleischer}}]{reichert_massively_2018}%
  \BibitemOpen
  \bibfield  {author} {\bibinfo {author} {\bibfnamefont {M.}~\bibnamefont
  {Reichert}}, \bibinfo {author} {\bibfnamefont {H.}~\bibnamefont {Defienne}},
  \ and\ \bibinfo {author} {\bibfnamefont {J.~W.}\ \bibnamefont {Fleischer}},\
  }\href@noop {} {\bibfield  {journal} {\bibinfo  {journal} {Scientific
  reports}\ }\textbf {\bibinfo {volume} {8}},\ \bibinfo {pages} {7925}
  (\bibinfo {year} {2018})}\BibitemShut {NoStop}%
\bibitem [{\citenamefont {Defienne}\ \emph
  {et~al.}(2018{\natexlab{a}})\citenamefont {Defienne}, \citenamefont
  {Reichert},\ and\ \citenamefont {Fleischer}}]{defienne_general_2018}%
  \BibitemOpen
  \bibfield  {author} {\bibinfo {author} {\bibfnamefont {H.}~\bibnamefont
  {Defienne}}, \bibinfo {author} {\bibfnamefont {M.}~\bibnamefont {Reichert}},
  \ and\ \bibinfo {author} {\bibfnamefont {J.~W.}\ \bibnamefont {Fleischer}},\
  }\href@noop {} {\bibfield  {journal} {\bibinfo  {journal} {Physical review
  letters}\ }\textbf {\bibinfo {volume} {120}},\ \bibinfo {pages} {203604}
  (\bibinfo {year} {2018}{\natexlab{a}})}\BibitemShut {NoStop}%
\bibitem [{\citenamefont {Moreau}\ \emph {et~al.}(2012)\citenamefont {Moreau},
  \citenamefont {Mougin-Sisini}, \citenamefont {Devaux},\ and\ \citenamefont
  {Lantz}}]{moreau_realization_2012}%
  \BibitemOpen
  \bibfield  {author} {\bibinfo {author} {\bibfnamefont {P.-A.}\ \bibnamefont
  {Moreau}}, \bibinfo {author} {\bibfnamefont {J.}~\bibnamefont
  {Mougin-Sisini}}, \bibinfo {author} {\bibfnamefont {F.}~\bibnamefont
  {Devaux}}, \ and\ \bibinfo {author} {\bibfnamefont {E.}~\bibnamefont
  {Lantz}},\ }\href@noop {} {\bibfield  {journal} {\bibinfo  {journal}
  {Physical Review A}\ }\textbf {\bibinfo {volume} {86}},\ \bibinfo {pages}
  {010101} (\bibinfo {year} {2012})}\BibitemShut {NoStop}%
\bibitem [{\citenamefont {Tasca}\ \emph {et~al.}(2012)\citenamefont {Tasca},
  \citenamefont {Izdebski}, \citenamefont {Buller}, \citenamefont {Leach},
  \citenamefont {Agnew}, \citenamefont {Padgett}, \citenamefont {Edgar},
  \citenamefont {Warburton},\ and\ \citenamefont {Boyd}}]{tasca_imaging_2012}%
  \BibitemOpen
  \bibfield  {author} {\bibinfo {author} {\bibfnamefont {D.~S.}\ \bibnamefont
  {Tasca}}, \bibinfo {author} {\bibfnamefont {F.}~\bibnamefont {Izdebski}},
  \bibinfo {author} {\bibfnamefont {G.~S.}\ \bibnamefont {Buller}}, \bibinfo
  {author} {\bibfnamefont {J.}~\bibnamefont {Leach}}, \bibinfo {author}
  {\bibfnamefont {M.}~\bibnamefont {Agnew}}, \bibinfo {author} {\bibfnamefont
  {M.~J.}\ \bibnamefont {Padgett}}, \bibinfo {author} {\bibfnamefont {M.~P.}\
  \bibnamefont {Edgar}}, \bibinfo {author} {\bibfnamefont {R.~E.}\ \bibnamefont
  {Warburton}}, \ and\ \bibinfo {author} {\bibfnamefont {R.~W.}\ \bibnamefont
  {Boyd}},\ }\href {\doibase 10.1038/ncomms1988} {\bibfield  {journal}
  {\bibinfo  {journal} {Nature Communications}\ }\textbf {\bibinfo {volume}
  {3}},\ \bibinfo {pages} {984} (\bibinfo {year} {2012})}\BibitemShut {NoStop}%
\bibitem [{sup()}]{supmat}%
  \BibitemOpen
  \href@noop {} {\bibinfo  {journal} {See Supplementary Material [url] for
  in-depth methods, additional results and theoretical demonstrations.}\
  }\BibitemShut {NoStop}%
\bibitem [{\citenamefont {Mandel}\ and\ \citenamefont
  {Wolf}(1965)}]{mandel_coherence_1965}%
  \BibitemOpen
\bibfield  {journal} {  }\bibfield  {author} {\bibinfo {author} {\bibfnamefont
  {L.}~\bibnamefont {Mandel}}\ and\ \bibinfo {author} {\bibfnamefont
  {E.}~\bibnamefont {Wolf}},\ }\href {\doibase 10.1103/RevModPhys.37.231}
  {\bibfield  {journal} {\bibinfo  {journal} {Reviews of Modern Physics}\
  }\textbf {\bibinfo {volume} {37}},\ \bibinfo {pages} {231} (\bibinfo {year}
  {1965})}\BibitemShut {NoStop}%
\bibitem [{\citenamefont {Chan}\ \emph {et~al.}(2007)\citenamefont {Chan},
  \citenamefont {Torres},\ and\ \citenamefont {Eberly}}]{chan_transverse_2007}%
  \BibitemOpen
  \bibfield  {author} {\bibinfo {author} {\bibfnamefont {K.~W.}\ \bibnamefont
  {Chan}}, \bibinfo {author} {\bibfnamefont {J.~P.}\ \bibnamefont {Torres}}, \
  and\ \bibinfo {author} {\bibfnamefont {J.~H.}\ \bibnamefont {Eberly}},\
  }\href {\doibase 10.1103/PhysRevA.75.050101} {\bibfield  {journal} {\bibinfo
  {journal} {Physical Review A}\ }\textbf {\bibinfo {volume} {75}},\ \bibinfo
  {pages} {050101} (\bibinfo {year} {2007})}\BibitemShut {NoStop}%
\bibitem [{\citenamefont {Law}\ and\ \citenamefont
  {Eberly}(2004)}]{law_analysis_2004}%
  \BibitemOpen
  \bibfield  {author} {\bibinfo {author} {\bibfnamefont {C.~K.}\ \bibnamefont
  {Law}}\ and\ \bibinfo {author} {\bibfnamefont {J.~H.}\ \bibnamefont
  {Eberly}},\ }\href {\doibase 10.1103/PhysRevLett.92.127903} {\bibfield
  {journal} {\bibinfo  {journal} {Physical Review Letters}\ }\textbf {\bibinfo
  {volume} {92}},\ \bibinfo {pages} {127903} (\bibinfo {year}
  {2004})}\BibitemShut {NoStop}%
\bibitem [{\citenamefont {Fedorov}(2015)}]{fedorov_schmidt_2015}%
  \BibitemOpen
  \bibfield  {author} {\bibinfo {author} {\bibfnamefont {M.~V.}\ \bibnamefont
  {Fedorov}},\ }\href {\doibase 10.1088/0031-8949/90/7/074048} {\bibfield
  {journal} {\bibinfo  {journal} {Physica Scripta}\ }\textbf {\bibinfo {volume}
  {90}},\ \bibinfo {pages} {074048} (\bibinfo {year} {2015})}\BibitemShut
  {NoStop}%
\bibitem [{\citenamefont {Defienne}\ \emph
  {et~al.}(2018{\natexlab{b}})\citenamefont {Defienne}, \citenamefont
  {Reichert},\ and\ \citenamefont {Fleischer}}]{defienne_adaptive_2018}%
  \BibitemOpen
  \bibfield  {author} {\bibinfo {author} {\bibfnamefont {H.}~\bibnamefont
  {Defienne}}, \bibinfo {author} {\bibfnamefont {M.}~\bibnamefont {Reichert}},
  \ and\ \bibinfo {author} {\bibfnamefont {J.~W.}\ \bibnamefont {Fleischer}},\
  }\href@noop {} {\bibfield  {journal} {\bibinfo  {journal} {Physical review
  letters}\ }\textbf {\bibinfo {volume} {121}},\ \bibinfo {pages} {233601}
  (\bibinfo {year} {2018}{\natexlab{b}})}\BibitemShut {NoStop}%
\bibitem [{\citenamefont {Peng}\ \emph {et~al.}(2018)\citenamefont {Peng},
  \citenamefont {Qiao}, \citenamefont {Xiang},\ and\ \citenamefont
  {Chen}}]{peng_manipulation_2018}%
  \BibitemOpen
  \bibfield  {author} {\bibinfo {author} {\bibfnamefont {Y.}~\bibnamefont
  {Peng}}, \bibinfo {author} {\bibfnamefont {Y.}~\bibnamefont {Qiao}}, \bibinfo
  {author} {\bibfnamefont {T.}~\bibnamefont {Xiang}}, \ and\ \bibinfo {author}
  {\bibfnamefont {X.}~\bibnamefont {Chen}},\ }\href {\doibase
  10.1364/OL.43.003985} {\bibfield  {journal} {\bibinfo  {journal} {Optics
  Letters}\ }\textbf {\bibinfo {volume} {43}},\ \bibinfo {pages} {3985}
  (\bibinfo {year} {2018})}\BibitemShut {NoStop}%
\bibitem [{\citenamefont {Qiu}\ and\ \citenamefont
  {She}(2012)}]{qiu_influence_2012}%
  \BibitemOpen
  \bibfield  {author} {\bibinfo {author} {\bibfnamefont {Y.}~\bibnamefont
  {Qiu}}\ and\ \bibinfo {author} {\bibfnamefont {W.}~\bibnamefont {She}},\
  }\href {\doibase 10.1007/s00340-012-5041-6} {\bibfield  {journal} {\bibinfo
  {journal} {Applied Physics B}\ }\textbf {\bibinfo {volume} {108}},\ \bibinfo
  {pages} {683} (\bibinfo {year} {2012})}\BibitemShut {NoStop}%
\bibitem [{\citenamefont {Hong}(2018)}]{hong_two-photon_2018}%
  \BibitemOpen
  \bibfield  {author} {\bibinfo {author} {\bibfnamefont {P.}~\bibnamefont
  {Hong}},\ }\href {\doibase 10.1063/1.5042504} {\bibfield  {journal} {\bibinfo
   {journal} {Applied Physics Letters}\ }\textbf {\bibinfo {volume} {113}},\
  \bibinfo {pages} {101109} (\bibinfo {year} {2018})}\BibitemShut {NoStop}%
\bibitem [{\citenamefont {Salter}\ \emph {et~al.}(2010)\citenamefont {Salter},
  \citenamefont {Stevenson}, \citenamefont {Farrer}, \citenamefont {Nicoll},
  \citenamefont {Ritchie},\ and\ \citenamefont
  {Shields}}]{salter_entangled-light-emitting_2010}%
  \BibitemOpen
  \bibfield  {author} {\bibinfo {author} {\bibfnamefont {C.~L.}\ \bibnamefont
  {Salter}}, \bibinfo {author} {\bibfnamefont {R.~M.}\ \bibnamefont
  {Stevenson}}, \bibinfo {author} {\bibfnamefont {I.}~\bibnamefont {Farrer}},
  \bibinfo {author} {\bibfnamefont {C.~A.}\ \bibnamefont {Nicoll}}, \bibinfo
  {author} {\bibfnamefont {D.~A.}\ \bibnamefont {Ritchie}}, \ and\ \bibinfo
  {author} {\bibfnamefont {A.~J.}\ \bibnamefont {Shields}},\ }\href {\doibase
  10.1038/nature09078} {\bibfield  {journal} {\bibinfo  {journal} {Nature}\
  }\textbf {\bibinfo {volume} {465}},\ \bibinfo {pages} {594} (\bibinfo {year}
  {2010})}\BibitemShut {NoStop}%
\end{thebibliography}%

\end{document}